\documentclass[prl,amsmath,amssymb,floatfix,superscriptaddress,notitlepage,showpacs,twocolumn]{revtex4}
\usepackage{graphicx}
\usepackage{hyperref}
\usepackage{bm}
\usepackage{amssymb}

\begin{document}

\title{Information-Entropic Stability Bound for Compact Objects: Application to Q-Balls and the Chandrasekhar Limit of Polytropes}

\author{Marcelo Gleiser}
\email{mgleiser@dartmouth.edu}
\affiliation{Department of Physics and Astronomy, Dartmouth College,
Hanover, NH 03755, USA}

\author{Damian Sowinski}
\email{Damian.Sowinski.GR@dartmouth.edu}
\affiliation{Department of Physics and Astronomy, Dartmouth College,
Hanover, NH 03755, USA}

\date{\today}

\begin{abstract}

Spatially-bound objects across diverse length and energy scales are characterized by a binding energy. We propose that their spatial structure is mathematically encoded as information in their momentum modes and described by a measure known as configurational entropy (CE) \cite{GS1}. Investigating solitonic $Q$-balls and stars with a polytropic equation of state $P=K\rho^{\gamma}$, we show that objects with large binding energy have low CE, whereas those at the brink of instability (zero binding energy) have near maximal CE. In particular, we use the CE to find the critical charge allowing for classically stable $Q$-balls and the Chandrasekhar limit for white dwarfs ($\gamma=4/3$) with an accuracy of a few percent.

\end{abstract}
\pacs{11.10.Lm, 03.65.Ge,04.40.Dg,05.45.Yv}
\maketitle

\section{Introduction}
From subatomic to astrophysical scales, spatially-bound objects result from the interplay between attractive and repulsive interactions whenever there is an energy gain. This behavior is well-illustrated when the object can be described as composed of one or more types of particles of mass $m_i$. The atomic nucleus is an obvious example, where $E_{\rm bind}=M-[Zm_p + (A-Z)m_n]$, with $M$ the nucleus mass, $A$ the mass number, and $m_{p(n)}$ the proton (neutron) mass. The instability of the nucleus under fission occurs when $E_{\rm bind}$ approaches $0$. In most classical and semi-classical applications, the main focus of this work, spatially-bound objects are solutions to the nonlinear equations modeling the system \cite{Solitons1,Solitons2} with energy density vanishing at spatial infinity. 

In most cases the methodology is similar: spatially-localized solutions are sought for certain boundary conditions; once found, their stability under certain classes of perturbations is explored, usually by varying one or more physical parameters. From solitons in field theory \cite{Solitons2,Rajamaran} to stellar objects \cite{Weinberg,Shapiro}, the onset of instability is usually identified by a growing perturbation.

In the present work we explore the physics of spatially-localized objects using a newly proposed quantity that, as will be shown here, discriminates between stable and unstable configurations {\it without} the use of perturbative techniques. Our method is based on the configurational entropy (CE) proposed by Gleiser and Stamatopoulos \cite{GS1}. As we will see, the CE identifies both the onset of instability of a spatially-bound configuration (the maximum in CE), as well as the approach towards the optimal, or most bound, state (the minimum in CE) with an accuracy of a few percent. 

We first introduce the definition of the configurational entropy. We then explore the stability of $Q$-balls, soliton-like objects constructed from a complex scalar field that owe their stability to a conserved $U(1)$ global charge $Q$ \cite{Coleman}. Next, we use the CE to investigate the stability of gravitationally-bound stars known as Newtonian polytropes \cite{Weinberg,Shapiro}, showing that the CE correctly predicts the onset of instability known as the Chandrasekhar limit for white dwarfs. We conclude with remarks on how to extend our approach to general-relativistic bound states and gravitational collapse.

\section{Configurational Entropy}
Since we are interested in structures with spatially-localized energy, consider the set of square-integrable bounded functions $f(\bf x) \in L^2({\bf R^d})$ and their Fourier transforms $F(\bf k)$. Now define the modal fraction $f(\bf k)$,
\begin{equation}
\label{modal fraction}
f(\mathbf{k}) = \frac{|F(\mathbf{k})|^2}{\int|F(\mathbf{k})|^2 d^d\bf{k}},
\end{equation}
where the integration is over all $\bf k$ where $F(\mathbf{k})$ is defined and $d$ is the number of spatial dimensions. $f(\bf{k})$ measures the relative weight of a given mode $\bf k$. For periodic functions where a Fourier series is defined, $f(\mathbf{k}) \rightarrow f_{\bf{n}}=|A_{\bf{n}}|^2/ \sum |A_{\bf{n}}|^2$, where $A_\mathbf{n}$ is the coefficient of the $\mathbf{n}$-th Fourier mode. (For details see \cite{GS1}.) In the continuum, we further introduce the normalized modal fraction, ${\tilde f}(\mathbf{k}) =f(\mathbf{k})/f(\mathbf{k}_{\rm max})$, where $\mathbf{k}_{\rm max}$ denotes the mode with maximum contribution to $f(\mathbf{k})$. The configurational entropy $S_C[{\tilde f}]$ is defined as 
\begin{equation}
\label{CE}
S_C[{\tilde f}] = - \int {\tilde f}(\mathbf{k})\ln [{\tilde f}(\mathbf{k})] \mathrm{d}^d \mathbf{k}.
\end{equation}
The integrand ${\tilde f}(\mathbf{k})\ln [{\tilde f}(\mathbf{k})]$ is the configurational entropy density. For configurations with a finite spatial extent (such as stars with radius $R$, see below) one must exclude irrelevant modes to avoid overcounting. In the spirit of Shannon's information entropy \cite{Shannon}, $S_C[{\tilde f}]$ gives an informational measure of the relative weights of different $k$-modes composing the configuration: in the $1$-dimensional discrete case it is maximized when all $N$ modes carry the same weight, the mode equipartition limit, $f(k_i)=1/N$ for any $k_i$, where $S_C[f]=\ln N$. If only a single mode is present, $S_C[f]=0$. $S_C[{\tilde f}]$ is, in a sense, an entropy of shape, a measure of the information content of a given spatial profile in terms of its momentum modes. The lower $S_C[{\tilde f}]$, the less information is needed to characterize the shape.
Our definition of configurational entropy should not be confused with that used in more traditional thermodynamic contexts, such as in protein folding \cite{Protein} and the liquid to glass transition \cite{Glass}.

\section{Q-balls}
$Q$-balls are nontopological solitons first proposed by Coleman \cite{Coleman}. Since then, they have been found in many model systems \cite{nonabelian,gaugeQ,susyQ,interacQ,Graham,cosmoQ}. In their simplest rendition (the one we will adopt here), they exist in models with a complex scalar field with a global $U(1)$ symmetry thus guaranteeing a conserved net charge $Q$. We use the model of Ref. \cite{GT} and refer the reader there for more details. The metric signature is $(+,-,\dots,-)$. The Lagrangian density is
\begin{equation}
{\cal L} = \partial_{\mu}\phi^{\dagger}\partial^{\mu}\phi -
m^2\phi^{\dagger}\phi + b(\phi^{\dagger}\phi)^2
-\frac{4c}{3}(\phi^{\dagger}\phi)^3~,
\label{lagrangian}
\end{equation}
where the constants $m^2$, $b$, and $c$ are real and positive. Writing the
field as $\phi({\bf x},t) = \frac{1}{\sqrt{2}}\Phi({\bf x})e^{i\omega t}$, and introducing the dimensionless
field $X^2 \equiv \sqrt{c/m^2}\Phi^2$, angular frequency $\omega' = \omega/m$,
and spacetime variables $x_{\mu}' = x_{\mu}m$, the mass-energy of a configuration is

\begin{eqnarray}
M[X] &=& \frac{m^{3-d}}{\sqrt{c}}\int \mathrm{d}^d\mathbf{x}'\left[\frac{1}{2}(\nabla^{'2} X)^2+\frac{1}{2}(1-\omega'^2)X^2 \right. \nonumber \\
& & \left. - \frac{b'}{4}X^4 + \frac{1}{6}X^6 \right] \equiv \frac{m^{3-d}}{\sqrt{c}}E~,
\label{energy}
\end{eqnarray}
where $b' \equiv b/(mc^{1/2})$. In this model, $Q$-balls exist for $2\leq b' \leq 4\sqrt{3}/3\simeq 2.309$. The lower bound guarantees that $\Phi_+$, the other minimum of $V(\Phi^2)$, exists, while the upper bound ensures that vacuum at $\Phi=0$ is the global minimum. $Q$-balls are thus nonperturbative excitations of the physical vacuum at $\Phi=0$. Henceforth we will drop the primes. (This means that $m\sqrt{c}=1$, as we see from definition of $b'$.) As $\omega\rightarrow 1$ we approach unstable configurations known as $Q$-clouds, characterized by small-amplitudes and large spatial extension \cite{Qclouds}. $Q$-balls are solutions of the equation
\begin{equation}
\nabla_d^2X = -\omega^2X +\frac{d V}{d X}
\equiv  U'(X)~,
\label{eqofmotion1}
\end{equation}
and hence live in the ``upside-down'' potential $-U(X)$. Solutions must satisfy the boundary conditions $X(0)=X_0$, $X'(0)=0$ and $X(r\rightarrow\infty)=0$, and are possible when $U(X_+) < 0$, which translates into the inequality $\omega_c \geq \sqrt{1-\frac{3 b'^2}{16}}$. Each solution leads to a conserved charge $Q = \omega \int \mathrm{d}^d\mathbf{x} \ \Phi^2$. ($Q$ is in units of $m^{2-d}/\sqrt{c}$.) For each solution we can compute the binding energy $E_{\rm bind} = M-Qm$, where $M$ is given in Eq. \ref{energy}. Using Eq. \ref{energy} and the dimensionless units condition $m\sqrt{c}=1$, we can rewrite the net binding energy in $d=3$ as, 
\begin{equation}
\label{Ebind}
\frac{E_{\rm bind}}{Qm} = \frac{E}{Q} - 1.
\end{equation}
Each choice of $b$ and $\omega$ corresponds to a $Q$-ball with binding energy given by Eq. \ref{Ebind}, a spherically-symmetric solution of Eq. \ref{eqofmotion1} with boundary conditions specified above. Classically stable configurations must have $E/Q < 1$. The solutions are found using a shooting method \cite{NumRec} on 64-bit floating point accuracy initial conditions, with a 4$^{th}$-order Runge-Kutta code using a step size of $.01$. 
 To compute the configurational entropy for $Q$-balls and other classically unstable configurations ($E/Q>1$) we use Eq. \ref{CE}, with $d=3$. 

\begin{figure}[htbp]
\includegraphics[width=\linewidth]{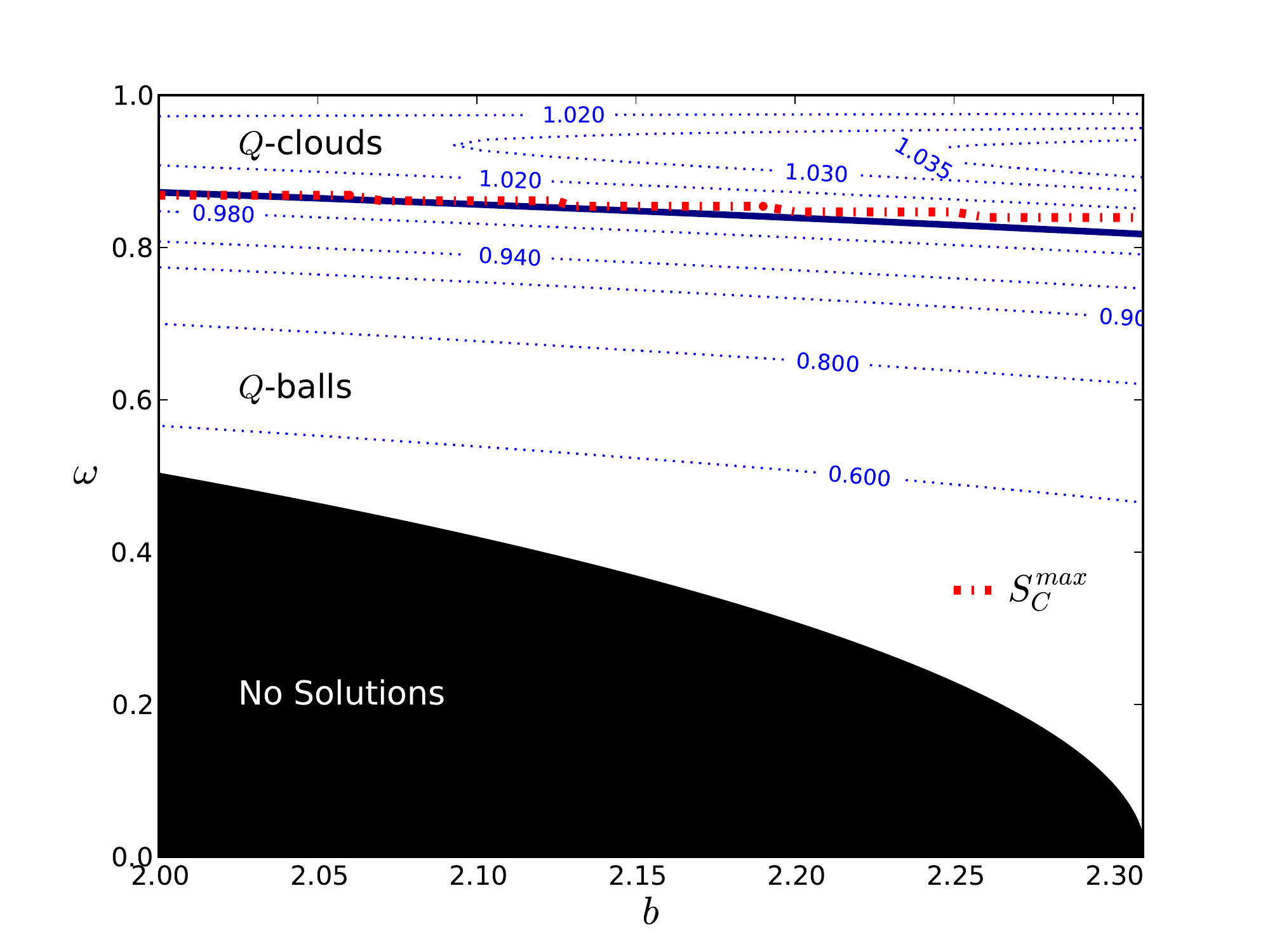}
\caption{(Color online.) The solution landscape of $Q$-balls for the potential of Eq. \ref{lagrangian}. The dotted (blue) lines represent contours of $E/Q$. The shaded region is forbidden by the inequality $\omega_c \geq \sqrt{1-3b^2/16}$, while the bold continuous line represents $E/Q=1$. The thick dotted (red) line is the maximum of the configurational entropy (cf. Figure \ref{E/Q}). Its near overlap with the $E/Q=1$ line confirms that the CE provides an efficient measure of $Q$-ball stability.}
\label{SolutionLandscape}
\end{figure}

In Figure \ref{SolutionLandscape} we plot the solution landscape of $Q$-balls as a function of $b$ and $\omega$. The dashed lines denote contours of constant ratio $E/Q$. $Q$-balls exist within the central region. The bold black line denotes the classical stability limit $E/Q=1$. We also indicate the $Q$-cloud region, within the classically unstable area above the $E/Q=1$ line.

\begin{figure}[htbp]
\includegraphics[width=\linewidth]{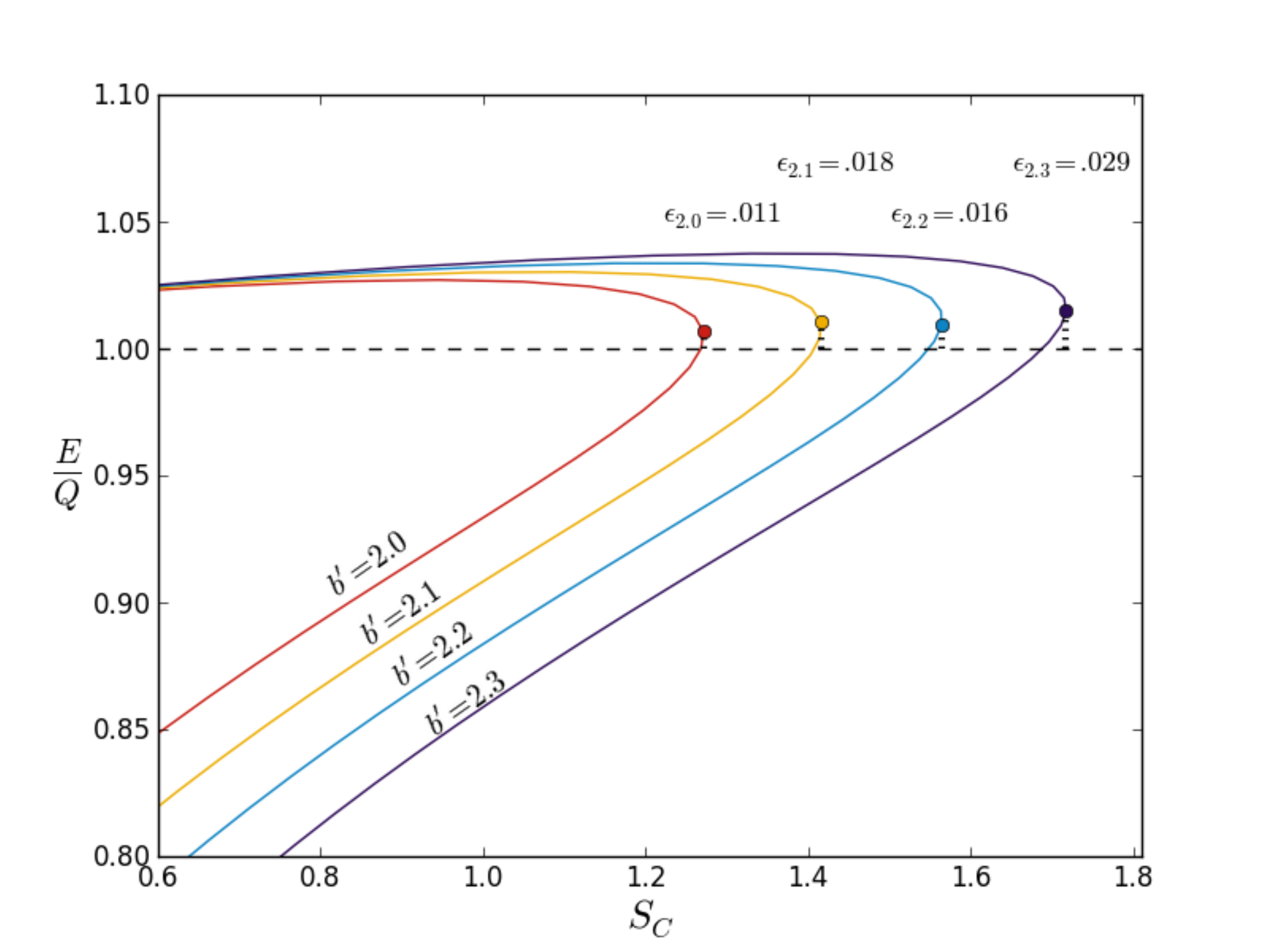}
\caption{$Q$-ball instability ratio versus configurational entropy parametrized by the value of $\omega$ for several values of $b$. Percent errors indicate the deviation of the maximum CE from the instability line (dashed).}
\label{E/Q}
\end{figure}

In Figure \ref{E/Q} we plot the ratio $E/Q$ of $Q$-balls for several allowed values of $b$ as a function of their configurational entropy. The CE is computed from the energy density $\rho(r)$ for each $Q$-ball solution. Each point corresponds to a solution of Eq. \ref{eqofmotion1}. The curves start at the lower left and run upwards with increasing $\omega$. $Q$-balls are classically stable when $E/Q<1$. There is a clear correlation between the binding energy and the configurational entropy: Figure \ref{SolutionLandscape} shows the maximal CE tracing the line of instability; Figure \ref{E/Q} shows that the maximum CE overshoots the line of instability by no more than $3\%$.

\section{Newtonian Stars and the Chandrasekhar Limit}
Next we investigate Newtonian polytropes and show how the CE can determine both the region of optimal stability and the region of instability, in particular the Chandrasekhar limit for relativistic white dwarf stars. In Newtonian gravity, stars are described as self-gravitating objects in hydrostatic equilibrium. For a spherically-symmetric, non-rotating configuration with pressure $P(r)$ and energy density $\rho(r)$ we have \cite{Weinberg,Shapiro},
\begin{equation}
\frac{d}{dr}\left[\frac{r^2}{\rho(r)}\frac{dP(r)}{dr}\right]=-4\pi Gr^2\rho(r).
\label{hydroeq}
\end{equation}
Eq. \ref{hydroeq} is supplemented by a general polytropic equation of state
\begin{equation}
P=K\rho^{\gamma},
\label{polyeq}
\end{equation}
where the constant $K$ depends on the entropy per nucleon and chemical composition. No heat flow throughout the object requires  $\gamma$ to be the adiabatic index, defined as the ratio of the heat capacities of the fluid at constant pressure and volume. Small mass, stable white dwarfs are well-modeled by $\gamma=5/3$ and $K=\frac{\hbar^2}{15m_e\pi^2}\left(\frac{3\pi^2}{m_N\mu}\right)^{5/3}$, where $m_{e(N)}$ is the electron (nucleon) mass, and $\mu\sim 2$ is the number of nucleons per electron. The largest mass white dwarfs are modeled by $\gamma=4/3$ and  $K= \frac{\hbar}{12\pi^2}\left(\frac{3\pi^2}{m_N\mu}\right)^{4/3}$\cite{Weinberg,Shapiro}. The binding energy for polytropes with $N$ nucleons, $E_{\rm bind}=M-Nm_N$, can be written as $E=-\frac{(3\gamma-4)}{(5\gamma-6)}\frac{GM^2}{R}$, where $M=\int_0^R 4\pi r^2\rho(r)dr$ is the star's mass and $R$ its radius, defined from $\rho(R)=0$. Solutions to Eqs. \ref{hydroeq} and \ref{polyeq} must satisfy $\rho(0)=\rho_c$ and $\rho'(0)=0$, and are found introducing new variables $\rho(r) = \rho_c\theta(\xi)^{1/(\gamma-1)}$ and $\xi=r/\alpha$, with $\alpha^2=\frac{K\gamma}{4\pi G(\gamma-1)}\rho_c^{(\gamma-2)}$. Equation \ref{hydroeq} then becomes the Lane-Emden equation with boundary conditions $\theta(0)=1$ and $\theta'(0)=0$,
\begin{equation}
\frac{1}{\xi^2}\frac{d}{d\xi}\xi^2\frac{d\theta}{d\xi} + \theta^{1/(\gamma-1)}=0.
\label{Lane-Emden}
\end{equation}
Solutions are found via a 4$^{th}$-order Runge-Kutta method with step size $10^{-3}$. The CE is computed from the energy density using Eq. \ref{CE}. Since stars have a finite radius (where $\rho(R)=0$ or, equivalently, $\theta(\xi_R)=0$), the integration is in the interval $k \in [k_{\rm min}=\pi/R,\infty)$. This ensures that only modes with wavelengths smaller than the polytrope contribute to the configurational entropy density.
 
\begin{figure}[htbp]
\includegraphics[width=\linewidth]{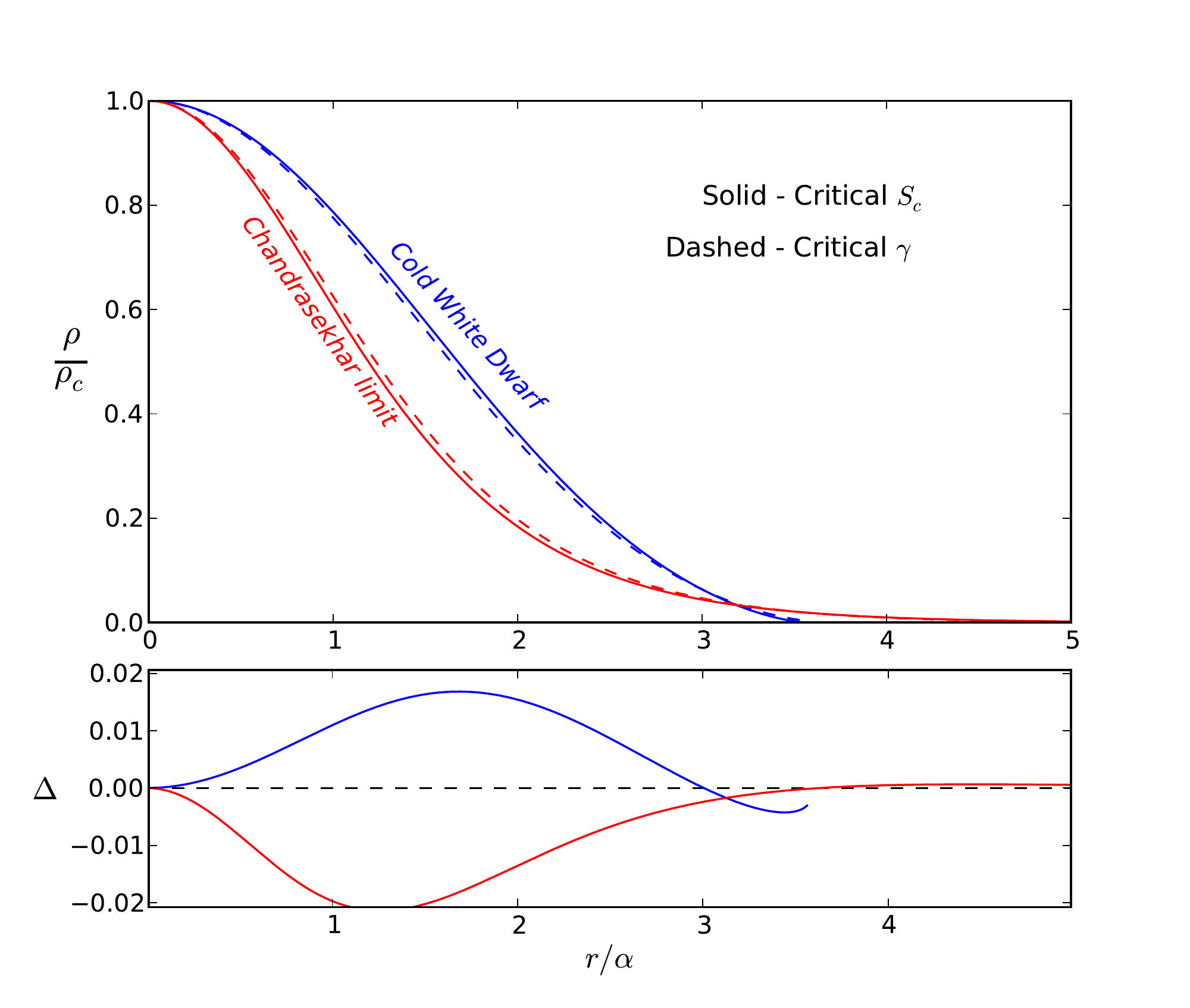}
\caption{Energy density vs. radius for polytropes with $\gamma=5/3$ (Cold White Dwarf) and $\gamma=4/3$ (Chandrasekhar limit). (Top) Dashed lines are solutions to the Lane-Emden equation, while continuous lines are solutions corresponding to the CE minimum and maximum, respectively. (Bottom) The difference between the two sets of curves.}
\label{whitedwarfs}
\end{figure}

\begin{figure}[htbp]
\includegraphics[width=\linewidth]{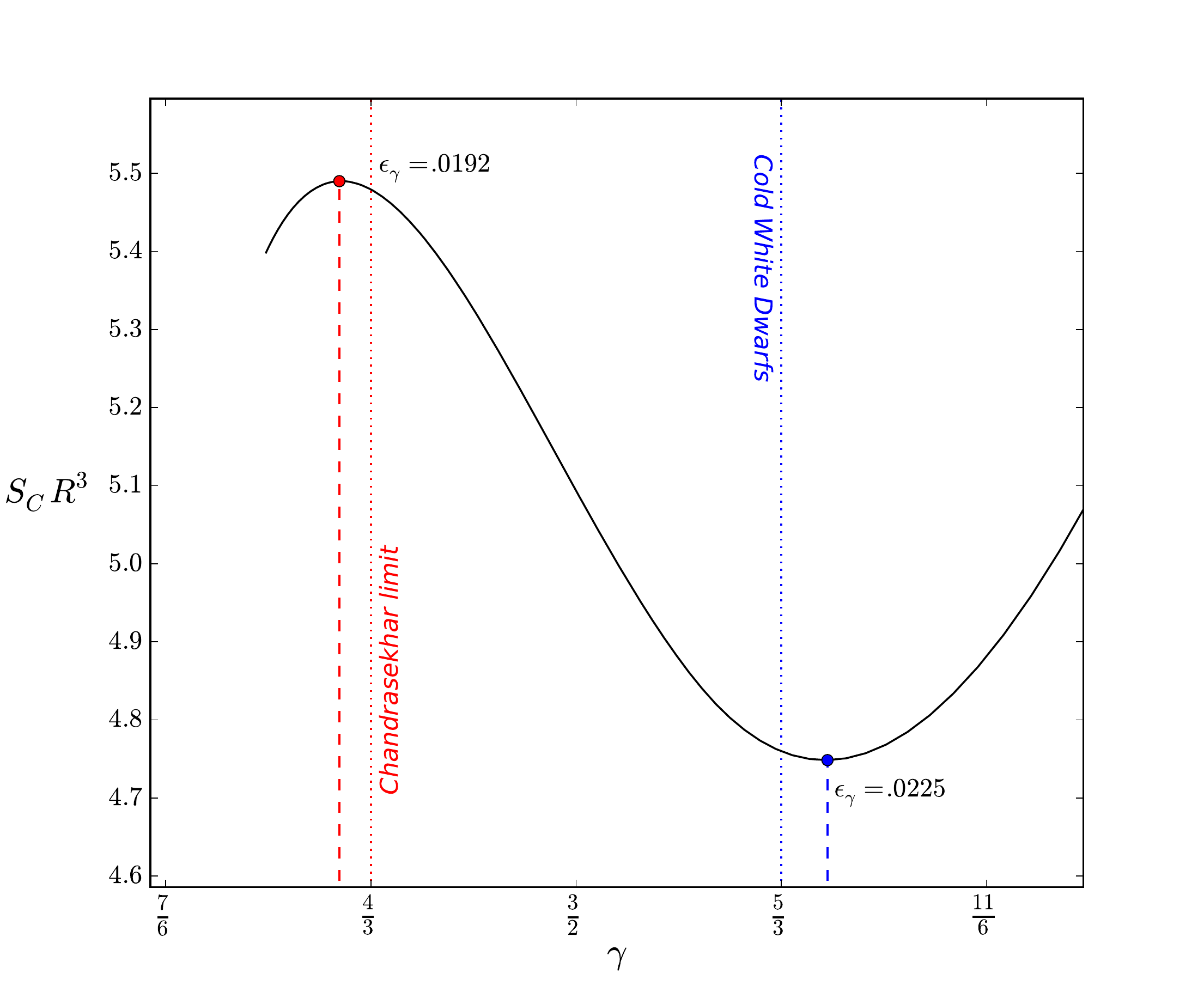}
\caption{Volume weighted configurational entropy for Newtonian polytropes. Note how the CE minimum is within a few percent of the stable polytrope with $\gamma=5/3$, modeling non-relativistic white dwarfs, while the CE maximum is within a few percent of the Chandrasekhar limit $\gamma=4/3$.}
\label{CEwhitedwarfs}
\end{figure}

In Figure \ref{whitedwarfs} (top) the dashed lines show the energy density profiles for polytropes with $\gamma=5/3$ (cold white dwarf) and $\gamma=4/3$ (Chandrasekhar limit). The solid lines correspond to the solutions for the minimum and maximum of the configurational entropy as depicted in Figure \ref{CEwhitedwarfs}. The bottom graph in Figure \ref{whitedwarfs} shows the difference ($\Delta$) between the two curves. The white dwarf with $\gamma=5/3$ corresponding to a non-relativistic stable bound state is well-approximated by the minimum of the CE, while the marginally stable ultra-relativistic $\gamma=4/3$ case is near the CE maximum. 

Recalling the results for $Q$-balls, we see that the configurational entropy provides a clear signature both for the optimal bound states (those with low CE) and for the marginally stable states (those with maximal CE). Indeed, we propose that the CE offers an independent criterion to determine the stability of spatially-bound configurations based solely in the informational content of their spatial profiles: the CE maximum represents the boundary between stability and instability.  


\section{Summary and Outlook}
We have investigated an entropic measure of ordering in field configuration space for nonlinear models with spatially-localized energy solutions. By studying the binding energy of $Q$-balls and of Newtonian polytropes as examples, we have found that this measure, the configurational entropy defined in Ref. \cite{GS1}, can be used to establish the region of stability for such compact objects with excellent accuracy. In particular, the CE maximum corresponds to the boundary between stability and instability, while optimally-bound structures have near-minimal CE. Although we have been unable so far to offer a formal proof relating the maximum of the configurational entropy to the boundary between stability and instability of spatially-bound objects, the evidence presented here, together we the results of Ref. \cite{GS1}, indicates that such proof is worth pursuing and that this relationship is quite general, possibly opening a new door in the study of complex systems in nature. It is possible that the agreement found here can be improved and made exact (within numerical accuracy) with a deeper understanding of the physical nature of the CE and its relation to the dynamical constraints of bound systems. Work along these lines is in progress.

We also intend to extend this study to general relativistic systems, exploring how the CE may give an indication of gravitational collapse such as in Oppenheimer-Snyder \cite{Weinberg,Shapiro} or in establishing the stability of boson stars \cite{bosonstars}. As in Ref. \cite{GS2}, we will then need a full time-dependent treatment. It is an open question whether there is a relation between the CE and Bekenstein's universal upper bound to entropy-to-energy ratio for bounded systems \cite{bekenstein}. 

MG is supported in part by a National Science Foundation grant PHY-1068027. MG and DS acknowledge support from the John Templeton Foundation grant under the New Frontiers in Astronomy \& Cosmology program.

\end{document}